\documentclass[aps, prl, superscriptaddress,amsmath,amssymb,twocolumn, longbibliography]{revtex4-2}
\usepackage{graphicx}
\usepackage{subfigure}
\usepackage{adjustbox}
\usepackage{bm}
\usepackage{color}
\usepackage{braket}
\usepackage{standalone}
\usepackage{multirow}
\usepackage{tikz}
\usepackage{mathrsfs}
\usepackage{dsfont}
\usepackage{comment}
\usepackage{amsmath,amssymb,bm}
\usepackage[colorlinks,bookmarks=true,citecolor=blue,linkcolor=red,urlcolor=blue]{hyperref}
\usepackage{cleveref}
\usepackage{orcidlink}
\usepackage{float}
\usepackage{subfiles}
\usepackage{commath}
\usepackage{balance}
\usepackage[normalem]{ulem}
\usepackage{balance}
\usepackage{pdfpages}
\makeatletter
\AtBeginDocument{\let\LS@rot\@undefined}
\makeatother

\newcommand{\bn}{{\mathbf n}}
\newcommand{\bs}{{\mathbf s}}

\newcommand{\he}{{\hat e}}

\newcommand{\bea}{\begin{eqnarray*}}
\newcommand{\eea}{\end{eqnarray*}}
\newcommand{\bean}{\begin{eqnarray}}
\newcommand{\eean}{\end{eqnarray}}

\begin{document}

\title{Magnetic and Lattice Ordered Fractional Quantum Hall Phases in Graphene}
\author{Jincheng An}
\email{Jincheng.An@uky.edu}
\affiliation{Department of Physics and Astronomy, University of Kentucky, Lexington, KY 40506, USA}
\author{Ajit C. Balram\orcidlink{0000-0002-8087-6015}}
\email{cb.ajit@gmail.com}
\affiliation{Institute of Mathematical Sciences, CIT Campus, Chennai, 600113, India}
\affiliation{Homi Bhabha National Institute, Training School Complex, Anushaktinagar, Mumbai 400094, India}
\author{Ganpathy Murthy\orcidlink{0000-0001-8047-6241}}
\email{murthy@g.uky.edu}
\affiliation{Department of Physics and Astronomy, University of Kentucky, Lexington, KY 40506, USA}	
\

\begin{abstract}
At and near charge neutrality, monolayer graphene in a perpendicular magnetic field is a quantum Hall ferromagnet. In addition to the highly symmetric Coulomb interaction, residual lattice-scale interactions, Zeeman, and sublattice couplings determine the fate of the ground state. Going beyond the simplest model with ultra-short-range residual couplings to more generic couplings, one finds integer phases that show the coexistence of magnetic and lattice order parameters. Here we show that fractional quantum Hall states in the vicinity of charge neutrality have even richer phase diagrams, with a plethora of phases with simultaneous magnetic and lattice symmetry breaking. 
\end{abstract}

\maketitle

{\bf \em Introduction.---}The integer \cite{IQHE_Discovery_1980, TKNN1982} and fractional \cite{FQHE_Discovery_1982, Laughlin_1983,PrangeGirvin1990, dassarma1996:qhe} quantum Hall effects  are the earliest known  topological insulators \cite{Kane_Mele_2DTI_PhysRevLett.95.146802,TIs_RevModPhys.82.3045}. 
The bulk has a charge gap, with the protected chiral edge modes \cite{Halperin_Edge_1982} being responsible for electrical transport. When internal degeneracies such as spin or valley are present, interactions lead to quantum Hall ferromagnetism \cite{QHFM_Fertig_1989, Shivaji_Skyrmion, QHFM_Yang_etal_1994, QHFM_Moon_etal_1995}. While the discovery of the integer/fractional quantum Hall effects (IQHEs/FQHEs) took place in semiconductor heterostructures, graphene \cite{Berger_etal_2004, Novoselov_etal_2004, Zhang_etal_2005,jiang2007:nu0, neto2009:rmp} emerged in the early 2000's as an alternate platform. In the absence of a perpendicular magnetic field $B_{\perp}$, monolayer graphene, with its honeycomb structure, has two Dirac band crossings at the corners of the Brillouin Zone, the $K$ and $K'$ points. For realistic $B_{\perp}$ fields, each Dirac point manifests a set of particle-hole symmetric Landau levels (LLs) with energies $E_n{\propto}{\rm sgn}(n)\sqrt{|n|B_{\perp}}$. The $n{=}0$ LL is special, with the wave function in each valley being restricted to one sublattice, a phenomenon known as valley-sublattice locking. The $n{=}0$ manifold of four (two for spin and two for valley) nearly degenerate LLs [called the zero-LLs (ZLLs)] has proven to be a fascinating example of quantum Hall ferromagnetism, one that is not yet fully understood. The charge neutral state, with two linear combinations of the four ZLLs filled, is labeled $\nu{=}0$. This state \cite{jiang2007:nu0,young2012:nu0, Maher_Kim_etal_2013,young2014:nu0, Magnon_transport_Yacoby_2018, Spin_transport_Lau2018, Magnon_Transport_Assouline_2021, Magnon_transport_Zhou_2022,li2019:stm, STM_Yazdani2021visualizing, STM_Coissard_2022}, and nearby fractional fillings have attracted a lot of attention experimentally \cite{Andrei_FQH2009, Bolotin_FQH2009, Hone_Multicomp_MGL2011, Hofstadter_Ashoori2013, GoldhaberGordon2015, Even_Den_Young2018, Young_Skyrmion_Solid_Graphene_2019, Huang21, Magnon_transport_Zhou_2022}. 

Based on a large body of work \cite{alicea2006:gqhe, KYang_SU4_Skyrmion_2006, Herbut1, Herbut2,abanin2006:nu0,brey2006:nu0}, a simple model Hamiltonian in the ZLLs was proposed by Kharitonov \cite{kharitonov2012:nu0}. The long-range Coulomb interaction is $SU(4)$ symmetric in the spin-valley space, and ultra-short-range (USR) residual interactions encode all the anisotropies. In the Hartree-Fock (HF) approximation, the ground state at $\nu{=}0$ breaks either the lattice symmetries or the $U(1)$ spin-rotation symmetry around the total field, but not both \cite{kharitonov2012:nu0}. Experimentally, there is strong evidence from scanning tunneling microscopy (STM) experiments \cite{li2019:stm, STM_Yazdani2021visualizing, STM_Coissard_2022, Yazdani_FQH_STM2023_1, Yazdani_FQH_STM2023_2} and magnon transmissions studies \cite{Magnon_transport_Yacoby_2018, Spin_transport_Lau2018, Magnon_Transport_Assouline_2021, Magnon_transport_Zhou_2022} that the ground states at $\nu{=}0$ and fractions around it spontaneously break both magnetic and lattice symmetries simultaneously. Relaxing the USR condition leads to ground states that do show the coexistence of spontaneous lattice and magnetic symmetry breaking \cite{Das_Kaul_Murthy_2022, De_etal_Murthy2022, Stefanidis_Sodemann2023}.

A variational approach based on USR anisotropic interactions was proposed by Sodemann and MacDonald for FQH states in the ZLLs \cite{Sodemann_MacDonald_2014}. For USR interactions, only the relative angular momentum $m{{=}}0$ Haldane pseudopotential $V_0$ \cite{Haldane_Pseudopot1983} is nonzero, which leads to a HF-like variational energy for the FQH state in the ZLLs \cite{Sodemann_MacDonald_2014}. Spontaneous breaking of lattice \emph{and} magnetic order seems to be extremely rare in this analysis or extensions thereof~\cite{Hegde_Sodemann2022}. In this Letter, we will explain the coexistence of lattice and magnetic orders in the most robust fractions in the vicinity of charge neutrality~\cite{li2019:stm, STM_Yazdani2021visualizing, STM_Coissard_2022, Yazdani_FQH_STM2023_1, Yazdani_FQH_STM2023_2, Magnon_transport_Yacoby_2018, Spin_transport_Lau2018, Magnon_Transport_Assouline_2021, Magnon_transport_Zhou_2022} by analyzing a more generic set of residual interactions that have a nonzero range. Specifically, we will show that when the residual anisotropic interactions have nonzero $V_0$ and $V_1$, a rich phase diagram with multiple regions of coexistence between spontaneously broken lattice and magnetic orders is obtained, especially in what is believed to be the physical regime of parameters. The introduction of nonzero-range interactions is motivated by physics. (i) From the renormalization group point of view, given the sizeable LL mixing, integrating out the other $|n|{>}0$ manifolds will inevitably generate nonzero range interactions \cite{Das_Kaul_Murthy_2022, De_etal_Murthy2022}. (ii) In Bernal-stacked bilayer graphene, which has many of the same phases as monolayer graphene~\cite{Huang21, Dora23}, trigonal warping can induce nonzero range effective interactions \cite{Murthy_BLG_2017, Khanna_BLG2023}. (iii) Projecting a USR interaction to a higher $|n|{\neq} 0$ Landau level manifold will also result in such interactions \cite{Stefanidis_Sodemann2022}. (iv) In the IQHE for $\nu{=}{\pm}1$, nonzero-range interactions are crucial in breaking unphysical degeneracies \cite{Lian_Rosch_Goerbig_2016, Lian_Goerbig_2017, Atteia_Goerbig_2021} and providing a spin-valley stiffness.

{\bf \em Hamiltonian and variational states.---}We include the Zeeman coupling $E_z$ and a sublattice symmetry breaking/valley Zeeman coupling $E_v$ induced by the moir\'e potential of the encapsulating hexagonal Boron Nitride (hBN)  \cite{KV_DGG_2013,KV_expt_2013,KV_hBN_Jung2015,KV_hBN_Jung2017}. In addition to the $SU(4)$ symmetric Coulomb interaction, a generic interacting Hamiltonian in the ZLLs has the following residual anisotropic interactions:
\begin{widetext}
\begin{equation}
H^{\rm an}=\frac{1}{2}\int d^2 {\bf r_1} d^2 {\bf r_2} \sum_{i, \alpha, \beta, \gamma, \delta} :\hat\psi^\dagger_\alpha(r_1)\tau_i^{\alpha\beta}\hat\psi_\beta(r_1)  
V_i( r_1- r_2)
\hat\psi^\dagger_\gamma(r_2)\tau_i^{\gamma\delta}\hat\psi_\delta(r_2):~,    
\end{equation}
\end{widetext}
where $i{=}x, y, z$, ${ V}_x{=}{ V}_y{=}{ V}_\perp(r_1{-}r_2)$, $\alpha, \beta, \gamma, \delta$ run over all the spin and valley indices, and the electron operators have been projected to the $n{=}0$ LL.
We abstractly define 
\begin{equation}
    {\cal V}_i=(g_{0i}{\cal P}_0+g_{1i}{\cal P}_1):{\hat \tau}_i{\hat \tau}_i:~, 
    \label{eq: pseudopot}
\end{equation}
where ${\cal P}_m$ represents a two-body interaction projected to the relative angular momentum $m$ state, and $g_{0i},g_{1i}$ are numbers with energy units. The Hartree and Fock couplings associated to these are $g_{i, H}{=}g_{0i}{+}g_{1i},~g_{i, F}{=}g_{0i}{-}g_{1i}$~\cite{SM}. It is widely believed~\cite{kharitonov2012:nu0,STM_Yazdani2021visualizing} that for real samples, $g_z{>}0,~g_{\perp}{<}0$, with their magnitudes not too different from each other.

The interaction terms have the symmetry $SU(2)_s{\otimes }U(1)_v{\otimes} Z_{2v}$ \cite{alicea2006:gqhe}, where the subscripts $s,v$ stand for spin and valley respectively. Upon adding $E_z,\ E_v$, the Hamiltonian has the symmetry $U(1)_s{\otimes} U(1)_v$. Upon the inclusion of 6-fermion terms, the $U(1)_v$ breaks down to a $Z_{3v}$. Thus, the spontaneous symmetry breaking of $U(1)_v$ (at the 4-fermion level) does not lead to Goldstone modes. We examine nonzero averages of the following one-body operators, written schematically as $\sigma_z$ [ferromagnet (FM)], $\tau_z\sigma_x$ [canted antiferromagnet (CAF)], $\tau_z$ [charge-density-wave (CDW)], $\tau_x$ [often termed Kekul\'e distorted (KD), but we use the more general term bond-ordered (BO)], $\tau_x\sigma_x$ [spin-valley entangled-X (SVEX)], $\tau_y\sigma_y$ (SVEY), and $\tau_z\sigma_z$ [antiferromagnet (AF)].

We focus upon the filling $\nu{=}{-}2{+}5/3$, which has been studied by variational \cite{Sodemann_MacDonald_2014} and exact diagonalization methods \cite{Papic_Goerbig_Regnault_Atypical_FQHE_MLG_1_3, Wu_Sodemann_Yasufumi_MacDonald_Jolicoeur_2014, Wu_Sodemann_MacDonald_Jolicoeur_SU3_SU4_2015, Le_Jolicoeur_SpinValley_FQH_MLG_2022}, albeit only for USR interactions. We will see how the introduction of $V_{1}$ immediately leads to states manifesting the simultaneous spontaneous breaking of magnetic and lattice symmetries. We use a generalization of a previously used ansatz \cite{Doucot_Goerbig_Skyrmion_2008, Sodemann_MacDonald_2014, Lian_Goerbig_2017, Goerbig_nu0_Skyrmion_Zoo_2021}. Any 4-component spinor can be written in a basis of direct products of 2-component spinors in the valley and spin spaces, denoted by $|\bn\rangle{\otimes}|\bs\rangle{\equiv} |\bn,\bs\rangle$. For example, for $\bs{=}\sin(\theta_s)[\cos(\phi_s)\he_x{+}\sin(\phi_s)\he_y]{+}\cos(\theta_s)\he_z$, $|\bs\rangle{=}[\cos(\theta_s/2),e^{i\phi_s}\sin(\theta_s/2)]^T$, and $|{-}\bs\rangle$ is obtained by $\theta{\rightarrow}\pi{-}\theta,~\phi{\rightarrow}\phi{+}\pi$ [details in the Supplemental Material (SM) \cite{SM}]. We can use the $U(1)$ symmetries to rotate $\bn,\bs$ so that they lie in the $xz$-plane, reducing the freedom to two angles $\theta_n,\theta_s{\in}[0,\pi]$, while $\phi_{n}$ and $\phi_{s}$ is either $0$ or $\pi$. We will need up to four spinors, either fully or partially occupied. We define
\begin{eqnarray}
    \label{eq:ansatz1}
    |f_1\rangle&=&\cos(\alpha_1/2)|\bn,\bs_a\rangle+\sin(\alpha_1/2)|-\bn,-\bs_b\rangle\nonumber\\
    |f_2\rangle&=&\cos(\alpha_2/2)|\bn,-\bs_a\rangle+\sin(\alpha_2/2)|-\bn,\bs_b\rangle\nonumber\\
    |f_3\rangle&=&\sin(\alpha_1/2)|\bn,\bs_a\rangle-\cos(\alpha_1/2)|-\bn,-\bs_b\rangle\nonumber\\
    |f_4\rangle&=&\sin(\alpha_2/2)|\bn,-\bs_a\rangle-\cos(\alpha_2/2)|-\bn,\bs_b\rangle,
\end{eqnarray}
where $\bs_a,\ \bs_b$ are, in general, different, and $\alpha_1,\alpha_2$ are introduced to cover a bigger range of variational states. 
We now have five continuously varying angles, $\theta_n,\theta_a,\theta_b,\alpha_1,\alpha_2$, and additionally the discrete possibilities $0,\pi$ for $\phi_{n}$ and $\phi_{s}$. This turns out to be sufficient to describe the most general two-component states (2CSs). For three-component states (3CSs), we need to take further linear combinations of the $|f_l\rangle$ to capture the full range of possibilities, as is detailed in the SM \cite{SM}.

{\it Variational states}: We use the composite fermion (CF) theory to construct variational FQHE states. According to the CF theory, the FQHE state at $\nu{=}n/(2n{\pm}1)$ maps into the $\nu^{*}{=}n$ IQHE state of CFs, which are electrons bound to two vortices~\cite{Jain89}. In the presence of (pseudo)spin, $n{=}n_{\uparrow}{+}n_{\downarrow}$, where $n_{\uparrow}$ ($n_{\downarrow}$) is the number of filled spin-up (spin-down) CF-LLs. This state, denoted by $[n_{\uparrow}, n_{\downarrow}]_{{\pm }2}$~\cite{Balram15}, is an eigenstate of the total spin and is described by the Jain wave function $\Psi^{[n_{\uparrow}, n_{\downarrow}]_{{\pm }2}}_{n/(2n{\pm} 1)}{=}\mathcal{P}_{\rm LLL}\Phi_{{\pm}n_\uparrow}\Phi_{{\pm}n_\downarrow}\prod_{j<k}(z_j{-}z_k)^{2}$, where the complex coordinate $z_j$ parametrizes the position of the $j$th electron, $\Phi_n$ is the Slater determinant wave function of $n$ filled LLs ($\Phi_{{-}|n|}{=}[\Phi_{|n|}]^*$), and $\mathcal{P}_{\rm LLL}$ denotes projection into the LLL. Unless otherwise stated, we shall use the ground states obtained from the exact diagonalization of the bare Coulomb interaction in the LLL to represent the CF states since for all accessible systems the two have a near unit overlap with each other~\cite{Ambrumenil88, Dev92a, Wu93, Jain07, Yang19a, Balram20b, Balram21b}. 

For $\nu{=}2/3$, we shall consider the fully polarized $[2,0]_{{-}2}$ and the singlet $[1,1]_{{-}2}$ states. The $[2,0]_{{-}2}$ state is nearly identical to the state obtained by hole conjugating the 1/3 Laughlin state~\cite{Balram16b, Balram21b} and since the latter can be treated semi-analytically, we shall use this representation for it. Using the fact that the 1/3 Laughlin state has $\langle V_{1} \rangle{=}0$ and the $\nu{=}1$ state has $\langle V_{1}/N_{\phi} \rangle{=}2$ in the thermodynamic limit, where $N_{\phi}$ is the number of flux quanta, one can show that the fully polarized 2/3 state has $\langle V_{1}/N_{\phi} \rangle{=}2/3$ in the thermodynamic limit~\cite{SM}. For the singlet state, by extrapolating finite system results in the spherical geometry~\cite{Haldane_Pseudopot1983} we numerically obtained $\langle V_{0}/N_{\phi} \rangle{=}0.001$ and $\langle V_{1}/N_{\phi} \rangle{=}0.49153$ in the thermodynamic limit~\cite{SM}. Furthermore, the singlet state has $\langle V^{\uparrow, \uparrow}_{1}\rangle{=}\langle V^{\uparrow, \downarrow}_{1}\rangle{=}\langle V^{\downarrow, \downarrow}_{1} \rangle$~\cite{SM}, where $V^{\sigma, \sigma'}_{m}$ is the $V_{m}$ pseudopotential energy for two electrons, one with spin $\sigma$ and the other with spin $\sigma'$. Since the singlet state has a very small $\langle V_{0}/N_{\phi}\rangle$ we can safely assume it to be hard-core (keeping a factor of $\prod_{j<k}(z_j{-}z_k)$ outside $\mathcal{P}_{\rm LLL}$ in $\Psi^{[1, 1]_{{-}2}}_{2/3}$ explicitly makes the state hard-core~\cite{Wu93}) since its small $V_{0}$ component will modify the phase boundaries only slightly. In the LLL, the singlet state has a lower bare Coulomb energy than the fully polarized state~\cite{Balram15a}. We do not consider the direct product of two 1/3 Laughlin states since that has a much higher Coulomb energy than the fully polarized and singlet 2/3 states~\cite{Faugno20}. 

{\bf \em Results.---}We will consider two candidate states for ${-}2{+}5/3$: $(1,[2,0]_{-2},0)$ and $(1,[1, 1]_{{-}2},0)$ (identical to $(1,2/3,0,0)$ and $(1,1/3,1/3,0)$ of Ref. \cite{Sodemann_MacDonald_2014} respectively). The first is a 2CS, while the second is a 3CS. The non-USR nature of the interactions will be quantified by the parameters $\Delta_{i}{=}2g_{1i}$, where $i{=}z,\perp$ and $g_{1i}$ is the strength of the $m{=}1$ pseudopotential appearing in Eq.~\eqref{eq: pseudopot}.

\begin{figure*}
    \includegraphics[width=1.0\textwidth,height=0.32\textwidth]{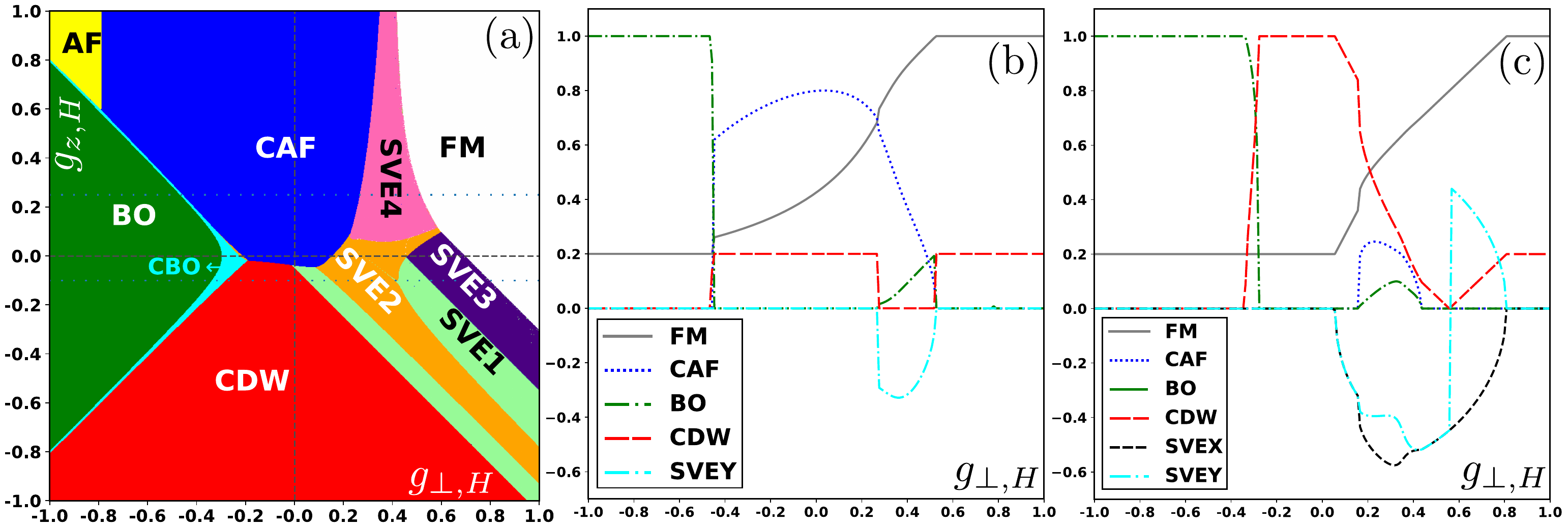}
    \caption{(a) Phase diagram for the the state $(1,[2,0]_{{-}2},0){\equiv}(1,2/3,0,0)$ at $E_z{=}1,\ E_v{=}0,\ \Delta_z{=}0.5,\ \Delta_\perp{=}0.7$. The SVE phases display a coexistence of magnetic and lattice symmetry breaking. (b) A section through the phase diagram for $g_{z, H}{=}0.25$ shows how the order parameters behave at the phase boundaries. Both first- and second-order transitions are seen. (c) A section at $g_{z, H}{=}{-}0.1$. All the transitions are second-order.
    }
    \label{fig: 1}
\end{figure*}

Fig.~\ref{fig: 1}(a) shows the phase diagram for the 2CS at $E_z{=}1,\ E_v{=}0$, with $\Delta_z{=}0.5,\ \Delta_{\perp}{=}0.7$, corresponding to a sample highly misaligned with the encapsulating hBN to suppress $E_v$ (or even a suspended sample) in a large tilted magnetic field $B$. In addition to the FM, BO, CDW, CAF, and AF phases known from the variational work with USR interactions \cite{Sodemann_MacDonald_2014, Hegde_Sodemann2022}, we have four additional phases with SVE order, and a tiny sliver of a canted BO (CBO) phase near the boundary between the CDW, CAF, and BO phases. As we show in the SM \cite{SM}, the SVE phase appears even in the USR limit at $E_v{=}0$. This result differs from that of Ref. \cite{Sodemann_MacDonald_2014}, because our variational ansatz [Eq.~\eqref{eq:ansatz1}] is more general. The four SVE phases differ in detail in the type of SVE order they display, and also in which other order parameters are nonvanishing.  In phase SVE1, ${\rm SVEX}{=}{\rm SVEY}$ and all other order parameters are nonzero. In phase SVE2, all order parameters are nonvanishing, and ${\rm SVEX}{\neq}{\rm SVEY}$. In phase SVE3, SVEX and BO orders vanish, but all other order parameters remain nonzero. Finally, in phase SVE4, CDW and SVEX order parameters vanish, but all others remain nonzero. Note that the spinors of both SVE1 and SVE3 are certain limits of the spinors of the KD-AF phase of Ref. \cite{Stefanidis_Sodemann2023}.
Fig.~\ref{fig: 1}(b) shows a cut through the phase diagram at $g_{z, H}{=}0.25$. The system is in the BO phase for $g_{\perp, H}{<}{-}1$. It crosses an extremely narrow region of the CBO phase near $g_{\perp, H}{=}{-}0.45$, one that is better resolved in Fig.~\ref{fig: 1}(c). We emphasize that this is a second-order transition. The system then makes a first-order transition to the CAF phase, which also has nonzero CDW order because of the unequal occupations of the spinors. Around $g_{\perp, H}{=}0.25$, the system makes a first-order transition into SVE4, which has CAF and BO order as well, but no CDW order. Near $g_{\perp, H}{=}0.5$ the system finally enters the FM phase via a first-order transition. Fig.~\ref{fig: 1}(c) shows a section at $g_{z, H}{=}{-}0.2$. All order parameters are continuous, implying that all the transitions are second-order. For $g_{z\perp}{<}{-}1$ the system is in the BO phase. As $g_{\perp, H}$ increases it crosses the narrow sliver of the CBO phase and enters the CDW phase, from which it then undergoes a transition to SVE1. After crossing SVE1, it enters SVE2, returns to SVE1, and then enters SVE3. Finally, it exits SVE3 and goes into the FM phase, which also has a nonzero CDW order parameter due to the unequal occupations of the two spinors. The spinors characterizing these phases are explored in the SM \cite{SM}.

\begin{figure*}
    \centering
    \includegraphics[width=1.0\textwidth,height=0.32\textwidth]{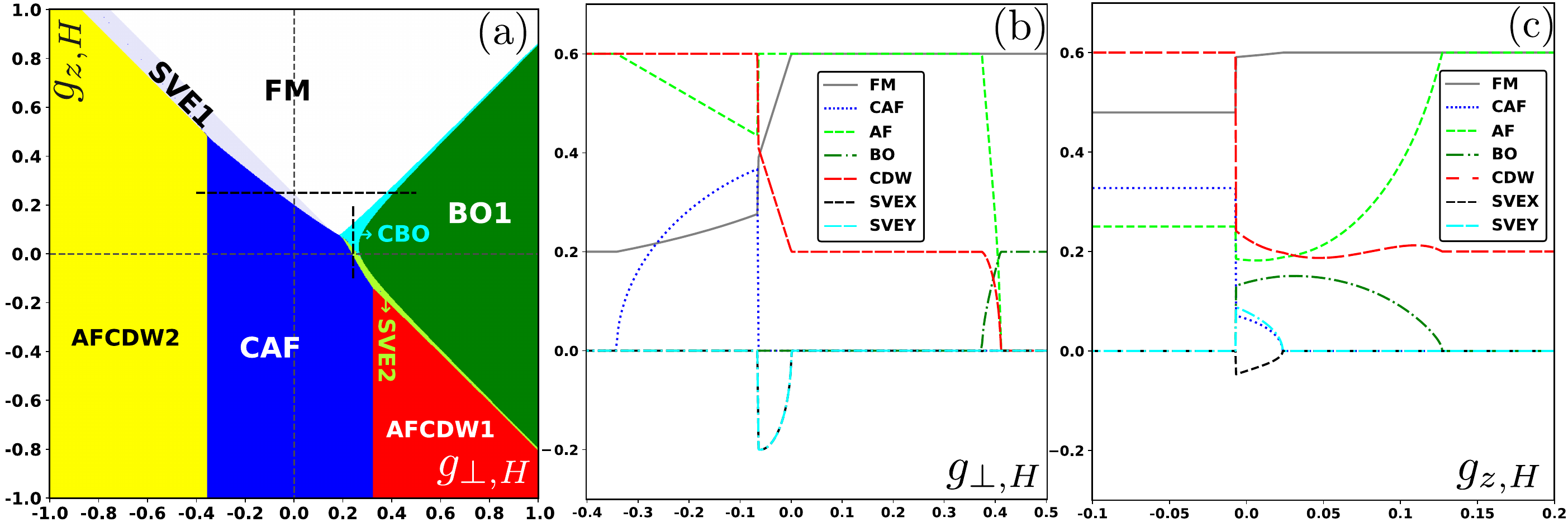}
    \caption{a) Phase diagram for the state $(1,[1,1]_{{-}2},0){\equiv}(1,1/3,1/3,0)$ at $E_z{=}1,\ E_v{=}0,\ \Delta_z{=}0.5,\ \Delta_{\perp}{=}0.7$. Eight different phases are seen, including two SVE phases manifesting a coexistence of lattice and magnetic order. (b) A horizontal section through the phase diagram at $g_{z, H}{=}0.25$. The system starts in the AFCDW2 phase, enters the CAF via a first-order transition, and then undergoes a second-order transition into the SVE1 phase. Next, it successively traverses the FM and CBO phases before making its way into the BO1 phase. All transitions except the first are second-order.  (c) A vertical section at $g_{z, \perp}{=}0.24$. The system starts in the CAF phase and enters the SVE2 phase via a first-order transition. It then traverses the CBO phase before finally going into the FM phase. The two later transitions are second-order.}
    \label{fig: 2}
\end{figure*}

Fig.~\ref{fig: 2}(a) shows the phase diagram of the 3CS for $E_z{=}1,\ E_v{=}0,\ \Delta_\perp{=}0.7,\ \Delta_z{=}0.5$. We see eight phases in all, as compared to the seven phases seen for USR interactions \cite{Sodemann_MacDonald_2014}. These phases are best described in terms of the spinors: $|f_1\rangle$ being fully filled, and $|f_2\rangle,\ |f_3\rangle$ spanning the space forming the singlet. One sees phases such as 
FM: $|K,\uparrow\rangle,\ |K^\prime,\uparrow\rangle,\ |K^\prime,\downarrow\rangle$, 
BO1: $|{-}\hat e_x,{\uparrow}\rangle$, $|\hat e_x,\uparrow\rangle$, $|\hat e_x,\downarrow\rangle$, 
AFCDW1: $|K,{\uparrow}\rangle$, $|K,{\downarrow}\rangle$, $|K^\prime,{\uparrow}\rangle$.
AFCDW2: $|K,\uparrow\rangle$, $|K,\downarrow\rangle$, $|K^\prime,\downarrow\rangle$,
CAF: $|K,-\bs_a\rangle,\ |K,\bs_a\rangle,\ |K^\prime,\bs_b\rangle$, 
CBO: $|-\bn_a,\uparrow\rangle,\ |\bn_a,\uparrow\rangle,\ |\bn_b,\downarrow\rangle $. 
SVE1 with $|K,\uparrow\rangle,\ |K^\prime,\downarrow\rangle, \cos\frac{\alpha}{2}|K,\downarrow\rangle+\sin\frac{\alpha}{2}|K^\prime,\uparrow\rangle$ will have SVEX${=}$SVEY order parameters.
SVE2 with all nonzero order parameters has no simple description and has to be found numerically~\cite{SM}.

Fig.~\ref{fig: 2}(b) shows a horizontal cut through the phase diagram at $g_{z, H}{=}0.25$. The system starts in AFCDW2, makes a second-order transition into CAF, a first-order transition into SVE1, and then transits the FM and CBO phases via second-order transitions to finally end in BO1. Fig.~\ref{fig: 2}(c) shows a vertical cut at $g_{\perp,H}{=}0.24$. The system starts in the CAF phase and enters SVE2 via a first-order transition, makes a second-order transition into the CBO phase, and finally goes into the FM phase. More details and other parameter choices, including $E_v{\neq}0$ for both types of states, are explored in the SM~\cite{SM}.

{\bf \em Conclusions and open questions.---}In this Letter, we have shown that given anisotropic interactions beyond USR, the filling $\nu{=}{-}2{+}5/3$ manifests phases that spontaneously break lattice and magnetic symmetries simultaneously. We examined a 2CS [conventionally denoted as $(1,2/3,0,0)$] and a 3CS [denoted $(1,1/3,1/3,0)$]. When sublattice symmetry is present ($E_v{=}0$), both types of states display rich phase diagrams with several coexistence phases that evolve with $E_v$~\cite{SM}. Strong field magnon transport experiments \cite{Magnon_transport_Zhou_2022} show that fractional states at nonzero $E_v$ develop magnetic order at a critical strength of $B_{\perp}$. Very recent STM experiments \cite{Yazdani_FQH_STM2023_1, Yazdani_FQH_STM2023_2} show lattice symmetry breaking even in FQH states. Though the two types of samples have differences, it seems likely that the spontaneous breaking of lattice and magnetic symmetries occurs simultaneously in all physical samples. Residual anisotropic interactions beyond USR produce such coexistence in large regions of the parameter space and are likely a key ingredient in the correct description of the physics. 

A host of open questions remain, such as, how does one uniquely identify a given state experimentally? Electrical and magnon transport \cite{Andrei_FQH2009, Bolotin_FQH2009, Hone_Multicomp_MGL2011, Hofstadter_Ashoori2013, GoldhaberGordon2015, Even_Den_Young2018, Young_Skyrmion_Solid_Graphene_2019, Huang21, Magnon_transport_Zhou_2022}, STM \cite{Andrei_STM2009,li2019:stm, STM_Yazdani2021visualizing,STM_Coissard_2022, Yazdani_FQH_STM2023_1, Yazdani_FQH_STM2023_2} including the imaging of order parameters near charged impurities \cite{STM_Yazdani2021visualizing}, and the detection of the gapless antiferromagnon \cite{JunZhu_2021} that characterizes any phase that spontaneously breaks the $U(1)$ spin-rotation symmetry are all important ingredients. Perhaps future spin-polarized STM experiments will be able to directly probe the spin-valley order. In another direction, as $B_\perp$ varies, all short-range couplings except $E_v$ are expected to scale like $B_\perp$~\cite{kharitonov2012:nu0}. The Coulomb interaction scales as $\sqrt{B_\perp}$. Thus, even within a particular type of state (2CS or 3CS) the system can undergo phase transitions as a function of $B_\perp$. Such transitions have indeed been seen in experiments \cite{Phase_transition_vs_Bperp_Yacoby2013, Maher_Kim_etal_2013, Magnon_transport_Zhou_2022, Huang21}, and their detailed understanding is an important task. Furthermore, due to the difference in Coulomb energy between the 2CSs and 3CSs, transitions may occur between them as $B_\perp$ varies~\cite{SM}. Extensions of our work include considering nonzero pseudopotentials for higher relative angular momenta or analyzing the phase diagrams of states such as ${-}2{+}7/5$ and even-denominator states (seen in graphene~\cite{Even_Den_Young2018, Kim19}). We are actively pursuing efforts in these directions~\cite{An_Balram_Murthy_future}. 

An important theoretical problem is to quantitatively understand how the residual anisotropic interactions become longer-ranged, assuming one starts from lattice scale interactions at high energies. This requires a systematic treatment of LL-mixing~\cite{Murthy_Shankar_LLmix, RG_Peterson_Nayak_2013, Peterson_Nayak_LLmix} coupled with a renormalization group approach \cite{Kang_Vafek_Moire_RG_2020, Chunli_FRG2023} involving all the couplings. We hope to address these and other important questions in the near future. 

We acknowledge useful discussions with Mark Goerbig and Thierry Jolicoeur. J.A. is grateful to  the University of Kentucky
Center for Computational Sciences and Information Technology Services Research Computing for the use of
the Morgan Compute Cluster. G.M. is grateful for the wonderful environment at the Aspen Center for Physics (NSF grant PHY-1607611). ACB and GM are grateful to the International Centre for Theoretical Sciences (ICTS) for supporting the program - Condensed Matter meets Quantum Information (code: ICTS/COMQUI2023/9), where parts of this project were conceived. Computational portions of this work were undertaken on the Nandadevi supercomputer, maintained and supported by the Institute of Mathematical Science's High-Performance Computing Center. Some of the numerical calculations were performed using the DiagHam libraries~\cite{DiagHam}. ACB thanks the Science and Engineering Research Board (SERB) of the Department of Science and Technology (DST) for funding support via the Mathematical Research Impact Centric Support (MATRICS) Grant No. MTR/2023/000002.

\bibliography{hall}

\newpage\newpage

\clearpage

\newpage\newpage



\section*{Supplementary material}

# Supplemental Material for "Magnetic and Lattice Ordered Fractional Quantum Hall Phases in Graphene"

Jincheng An,[1, *] Ajit C. Balram,[2, 3, †] and Ganpathy Murthy[1, ‡]

[1]*Department of Physics and Astronomy, University of Kentucky, Lexington, KY 40506, USA*
[2]*Institute of Mathematical Sciences, CIT Campus, Chennai, 600113, India*
[3]*Homi Bhabha National Institute, Training School Complex, Anushaktinagar, Mumbai 400094, India*

This supplemental material contains many technical details about the results presented in the main text. In Sec. SI we briefly review the Haldane pseudopotentials and the matrix elements of the two-body interactions restricted to relative angular momentum $m=0,1$. In Sec. SII we present the rationale behind the structure of the spinors we have chosen. The calculation of the variational energy for general fillings is presented in Sec. SIII. In Sec. SIV we present the details of the phases shown in the main text for the state $(1, [2,0]_{-2}, 0) \equiv (1, 2/3, 0, 0)$. Since we focused on the valley symmetric situation ($E_v=0$) in the main text, we also present some phase diagrams showing the evolution with $E_v$. In Sec. SV we do the same for the state $(1, [1,1]_{-2}, 0) \equiv (1, 1/3, 1/3, 0)$. Finally, we consider the Coulomb interaction as well, taking into account the Coulomb energy difference between the two-component states (2CS) and three-component states (3CS), and show a few phase diagrams that manifest transitions between the 2CS and 3CS in Sec. SVI.

## SI. HALDANE PSEUDOPOTENTIALS

The Haldane pseudopotentials [1] is a compact way of parameterizing any two-body interaction that is translationally and rotationally invariant, projected to a single Landau level. The key idea is that such an interaction depends only on the magnitude of the relative coordinate between two particles and therefore can be expanded in the basis of the center of mass angular momentum $M$ and the relative angular momentum $m$. The four-fermion interaction matrix element in the symmetric gauge can be written as

$$V_{m_1 m_2 m_3 m_4} = \langle m_1, m_2 | \hat{V} | m_3, m_4 \rangle = \sum_m V_m \sum_M \langle m_1, m_2 | M, m \rangle \langle M, m | m_3, m_4 \rangle, \tag{S1}$$

where the $V_m$ are the Haldane pseudopotentials [1]. Since we keep only $V_0$ and $V_1$, we need

$$\langle m_1, m_2 | m_1+m_2, 0 \rangle = \frac{1}{\sqrt{2^{m_1+m_2}}} \sqrt{\frac{(m_1+m_2)!}{m_1! m_2!}}, \quad \langle m_1, m_2 | m_1+m_2-1, 1 \rangle = \frac{m_1 - m_2}{\sqrt{2^{m_1+m_2}}} \sqrt{\frac{(m_1+m_2-1)!}{m_1! m_2!}}. \tag{S2}$$

If one carries a Hartree-Fock (HF) approximation on a state with well-defined total angular momentum,

$$(V_0^H)_{m_1 m_2} = (V_0)_{m_1 m_2 m_2 m_1} = (V_0)_{m_1 m_2 m_1 m_2} = (V_0^F)_{m_1 m_2},$$
$$(V_1^H)_{m_1 m_2} = (V_1)_{m_1 m_2 m_2 m_1} = -(V_1)_{m_1 m_2 m_1 m_2} = -(V_1^F)_{m_1 m_2}, \tag{S3}$$

the Hartree coupling will be identical to the Fock coupling for $\hat{V}_0$ while their sign will be opposite for $\hat{V}_1$. For future convenience, we define two sums that appear repeatedly in the energy functional and happen to both be equal to 2.

$$S_0 = \frac{1}{N_\phi} \sum_{m_1 m_2} (V_0)_{m_1 m_2 m_2 m_1} = S_1 = \frac{1}{N_\phi} \sum_{m_1 m_2} (V_1)_{m_1 m_2 m_2 m_1} = 2. \tag{S4}$$

## SII. SPINOR ANSATZ FOR VARIATION

Our ansatz for the spinors is a generalization of ansatzes that have been used in the literature [2–7]. We begin with some general considerations. We use the $U(1)_s \otimes U(1)_v$ symmetry of the Hamiltonian to make all the spinors real, which means that they all live on the three-sphere $S^3$. To see what this restriction means, let us consider a product

---

* Jincheng.An@uky.edu
† cb.ajit@gmail.com
‡ murthy@g.uky.edu

spinor $|\mathbf{n}, \mathbf{s}\rangle = |\mathbf{n}\rangle \otimes |\mathbf{s}\rangle$, where $|\mathbf{n}\rangle = [\cos(\theta_n/2), e^{i\phi_n}\sin(\theta_n/2)]^T$, $|\mathbf{s}\rangle = [\cos(\theta_s/2), e^{i\phi_s}\sin(\theta_s/2)]^T$. The imaginary parts appear via the projection of $\mathbf{n}$ or $\mathbf{s}$ in their respective $xy$-planes. So we assume that if the occupied spinors have nontrivial projections in the spin $xy$-plane, say, they will all align so that a single $U(1)_s \otimes U(1)_v$ rotation will make them real. One spinor can thus be described by three angles. Two orthogonal spinors need five angles (three for each spinor, with one orthogonality constraint), and three need six angles to describe. We also need ansatzes that can describe states where each spinor can have a different spin direction or a different valley direction. Thus, we start with the following ansatz

$$|f_1\rangle = \cos(\alpha_1/2)|\mathbf{n}, \mathbf{s}_a\rangle + \sin(\alpha_1/2)|-\mathbf{n}, -\mathbf{s}_b\rangle, \quad |f_2\rangle = \cos(\alpha_2/2)|\mathbf{n}, -\mathbf{s}_a\rangle + \sin(\alpha_2/2)|-\mathbf{n}, \mathbf{s}_b\rangle,$$
$$|f_3\rangle = \sin(\alpha_1/2)|\mathbf{n}, \mathbf{s}_a\rangle - \cos(\alpha_1/2)|-\mathbf{n}, -\mathbf{s}_b\rangle, \quad |f_4\rangle = \sin(\alpha_2/2)|\mathbf{n}, -\mathbf{s}_a\rangle - \cos(\alpha_2/2)|-\mathbf{n}, \mathbf{s}_b\rangle. \tag{S5}$$

We use the $U(1)_s \otimes U(1)_v$ symmetry to restrict the azimuthal angles $\phi_i$ to be either $0$ or $\pi$. One can see that the description of $|f_i\rangle$ requires only five continuously varying angles, namely $\alpha_1$, $\alpha_2$, $\theta_n$, $\theta_a$, $\theta_b$. Thus, this ansatz will be sufficient to describe two orthogonal spinors but is not general enough to describe three orthogonal spinors by our earlier counting argument. Hence, in describing states such as $(1, [1, 1]_{-2}, 0) \equiv (1, 1/3, 1/3, 0)$, where three orthogonal spinors are necessarily involved, we make a further set of rotations by introducing two more angles $\eta_1$, $\eta_2$ as shown below.

$$|\tilde{f}_1\rangle = \cos(\eta_1/2)|f_1\rangle + \sin(\eta_1/2)|f_2\rangle, \quad |\tilde{f}_2\rangle = -\sin(\eta_1/2)|f_1\rangle + \cos(\eta_1/2)|f_2\rangle,$$
$$|\tilde{f}_3\rangle = \cos(\eta_2/2)|f_3\rangle + \sin(\eta_2/2)|f_4\rangle, \quad |\tilde{f}_4\rangle = -\sin(\eta_2/2)|f_3\rangle + \cos(\eta_2/2)|f_4\rangle, \tag{S6}$$

Our description now uses seven continuously varying angles and is therefore slightly redundant. However, we believe it is complete. We have found through experience that the first ansatz, Eq. (S5), misses certain phases for the state $(1, [1, 1]_{-2}, 0)$ with fillings $(1, 1/3, 1/3, 0)$, while the second sufficient but redundant ansatz, Eq. (S6), does capture them.

For $\vec{\nu} = (\nu_1, \nu_2, \cdots)$, all order parameters $\hat{O}'s$ mentioned in this work are computed as $\langle \hat{O} \rangle = \sum_i \nu_i \text{Tr}(P_i \hat{O})/\sum_i \nu_i$, where $P_i = |f_i\rangle\langle f_i|$ is the project onto the state $|f_i\rangle$. Phases are differentiated by numerically determining whether various order parameters $\langle \hat{O} \rangle$'s are zero or not.

## SIII. VARIATIONAL ENERGY

With guiding center index sums being implicit, the residual anisotropic interaction we consider, which has support only in the relative angular momentum $m=0, 1$ channels, can be written as

$$\hat{H}^{\text{an}} \propto g_\perp : \hat{\tau}^x \hat{\tau}^x + \hat{\tau}^y \hat{\tau}^y : + g_z : \hat{\tau}^z \hat{\tau}^z :,$$
$$g_\perp = g_{0,\perp} + g_{1,\perp}, \quad g_z = g_{0,z} + g_{1,z}, \quad \hat{\tau}^i = c_\alpha^\dagger \tau_{\alpha\beta}^i c_\beta. \tag{S7}$$

Due to Eq. (S3), $g_{0,\perp}^H = g_{0,\perp}^F$, $g_{1,\perp}^H = -g_{1,\perp}^F$; $g_{0,z}^H = g_{0,z}^F$, $g_{1,z}^H = -g_{1,z}^F$. We will parameterize the non-USR nature of the interactions using the differences between Hartree and Fock couplings by
$\Delta_\perp = g_\perp^H - g_\perp^F = 2g_{1,\perp}$, $\Delta_z = g_z^H - g_z^F = 2g_{1,z}$. The Hamiltonian for the residual anisotropic interactions written out in full is

$$\hat{H}^{\text{an}} = \frac{1}{2} \sum_{\{m_i\}, \alpha\beta\gamma\delta} V_{m_1 m_2 m_3 m_4}^{\alpha\beta\gamma\delta} c_{m_1\alpha}^\dagger c_{m_2\beta}^\dagger c_{m_3\gamma} c_{m_4\delta}. \tag{S8}$$

It is crucial that the matrix elements have a product structure, with a part that depends only on the guiding center indices $m_i$ multiplied by a part that depends on the spin-valley indices $\alpha\beta\gamma\delta$. Explicitly,

$$V_{m_1 m_2 m_3 m_4}^{\alpha\beta\gamma\delta} = \underline{\tau}_{\alpha\delta}^z \underline{\tau}_{\beta\gamma}^z \bigg( g_{0,z}(V_0)_{m_1 m_2 m_3 m_4} + g_{1,z}(V_1)_{m_1 m_2 m_3 m_4} \bigg)$$
$$+ \big(\underline{\tau}_{\alpha\delta}^x \underline{\tau}_{\beta\gamma}^x + \underline{\tau}_{\alpha\delta}^y \underline{\tau}_{\beta\gamma}^y\big) \big(g_{0,\perp}(V_0)_{m_1 m_2 m_3 m_4} + g_{1,\perp}(V_1)_{m_1 m_2 m_3 m_4}\big), \tag{S9}$$

where $\underline{\tau}$ are the 4×4 matrices in spin-valley space with the spin-part set to identity and the valley part being the 2×2 Pauli matrix $\tau$. In computing the variational energy, since the initial and final states are identical, either all four Fermi operators operate on the same spinor, or a creation-destruction pair acts on one spinor, while the other creation-destruction pair acts on another spinor [3]. When at least two of the fermion operators belong to a fully

occupied spinor we can make a further simplification, that of replacing the average of the product of four fermion operators by the product of averages of $\langle c^\dagger c \rangle$.

$$\langle \hat{H}^{\mathrm{an}} \rangle = \frac{N_\phi}{2} \sum_{m=0,1} \nu_1 \nu_2 S_m \mathcal{E}^{\mathrm{an}}(P_1, P_2, g_{m,\perp}, g_{m,z})$$

$$\mathcal{E}^{\mathrm{an}}(P_1, P_2, g_\perp, g_z) = g_\perp^H \big( \mathrm{Tr}(P_1 \tau_x) \mathrm{Tr}(P_2 \tau_x) + \mathrm{Tr}(P_1 \tau_y) \mathrm{Tr}(P_2 \tau_y) \big) + g_z^H \mathrm{Tr}(P_1 \tau_z) \mathrm{Tr}(P_2 \tau_z)$$
$$- g_\perp^F \big( \mathrm{Tr}(P_1 \tau_x P_2 \tau_x) + \mathrm{Tr}(P_1 \tau_y P_2 \tau_y) \big) - g_z^F \mathrm{Tr}(P_1 \tau_z P_2 \tau_z) \tag{S10}$$

First, we consider the variational energy when a certain number of spinors are fully filled, and one spinor is $2/3$ filled. We will label this as $(1^n, 2/3) \equiv |1 f_1, \cdots, 1 f_n, (2/3) f_{n+1}\rangle$. It is convenient to define a separate projector $(P_I = \sum_{i=1}^n P_i)$ for the fully filled levels. The variational energy of $(1^n, 2/3)$ is given by

$$E^{\mathrm{an}}_{(1^n, \frac{2}{3})} = \frac{1}{2} \sum_{m=0,1} S_m \mathcal{E}^{\mathrm{an}}(P_I, P_I, g_{m,\perp}, g_{m,z}) + \frac{2}{3} \sum_{m=0,1} S_m \mathcal{E}^{\mathrm{an}}(P_I, P_{n+1}, g_{m,\perp}, g_{m,z})$$
$$+ \frac{1}{2} \frac{\langle 2/3 | \hat{V}_1 | 2/3 \rangle}{2 N_\phi} \mathcal{E}^{\mathrm{an}}(P_{n+1}, P_{n+1}, g_{1,\perp}, g_{1,z}) - \mathrm{Tr}\big(P(E_z \sigma_z + E_v \tau_z)\big). \tag{S11}$$

By an antiunitary particle-hole transformation [3] we obtain $\langle 2/3 | \hat{V}_1 | 2/3 \rangle = \langle 1 | \hat{V}_1 | 1 \rangle / 3$.

Now we turn to the filling $(1^n, \frac{1}{3}, \frac{1}{3})$ where there are $n$ integer filled spinors and two $\frac{1}{3}$ fractional filled spinors $|\Psi_{(1^n, 1/3, 1/3)}\rangle = |1 f_1, .., 1 f_n, (1/3) f_{n+1}, (1/3) f_{n+2}\rangle$. It is understood that the fractionally filled spinors are in the singlet state $(1, [1,1]_{-2}, 0)$ as described in the main text, and are not a product of two $1/3$ Laughlin states. The energy functional is

$$E^{\mathrm{an}}_{(1^n, \frac{1}{3}, \frac{1}{3})} = \frac{1}{2} \sum_{m=0,1} S_m \mathcal{E}^{\mathrm{an}}(P_I, P_I, g_{m,\perp}, g_{m,z}) + \frac{1}{3} \sum_{m=0,1} S_m \mathcal{E}^{\mathrm{an}}(P_I, P_{n+1} + P_{n+2}, g_{m,\perp}, g_{m,z})$$
$$+ \frac{a}{4} \mathcal{E}^{\mathrm{an}}(P_{n+1} + P_{n+2}, P_{n+1} + P_{n+2}, g_{1,\perp}, g_{1,z}) - \mathrm{Tr}\big((P(E_z \sigma_z + E_v \tau_z)\big) \tag{S12}$$

The third term on the RHS, the interaction in the correlated fractionally filled spinors is derived as follows. We project to that two-dimensional (2D) subspace and write all matrices such as the $\tau^\mu$ in that basis as $\tilde{\tau}^\mu_{ij} = \langle f_i | \tau^\mu | f_j \rangle$, with $i, j = 1, 2$ representing the states $|f_{n+1}\rangle$, $|f_{n+2}\rangle$ respectively. Listing out these possibilities, specializing to the $m=1$ angular momentum channel, and using antisymmetry we obtain

$$\langle \Psi | \frac{1}{2} \sum_{\{m_i\} j_1 j_2 j_3 j_4} g_\mu V_{m_1 m_2 m_3 m_4} \tilde{\tau}^\mu_{j_1 j_4} \tilde{\tau}^\mu_{j_2 j_3} c^\dagger_{j_1 m_1} c^\dagger_{j_2 m_2} c_{j_3 m_3} c_{j_4 m_4} | \Psi \rangle$$
$$= \frac{g_\mu}{2} \sum_{\{m_i\}} (V_1)_{m_1 m_2 m_3 m_4} \bigg( \tilde{\tau}^\mu_{11} \tilde{\tau}^\mu_{11} \langle \Psi | c^\dagger_{1 m_1} c^\dagger_{1 m_2} c_{1 m_3} c_{1 m_4} | \Psi \rangle + \tilde{\tau}^\mu_{22} \tilde{\tau}^\mu_{22} \langle \Psi | c^\dagger_{2 m_1} c^\dagger_{2 m_2} c_{2 m_3} c_{2 m_4} | \Psi \rangle$$
$$+ 2(\tilde{\tau}^\mu_{11} \tilde{\tau}^\mu_{22} + \tilde{\tau}^\mu_{12} \tilde{\tau}^\mu_{21}) \langle \Psi | c^\dagger_{1 m_1} c^\dagger_{2 m_2} c_{2 m_3} c_{1 m_4} | \Psi \rangle \bigg). \tag{S13}$$

One can no longer claim that $(m_3, m_4)$ is a permutation of $(m_1, m_2)$ because the state is strongly correlated. The averages have to be computed by exact diagonalization on the $[1,1]_{-2}$ singlet state. Leaving all the $m$ sums implicit, the result is the following:

$$(1/N_\phi) \langle \Psi | V_1 c^\dagger_1 c^\dagger_1 c_1 c_1 | \Psi \rangle = (1/N_\phi) \langle \Psi | V_1 c^\dagger_2 c^\dagger_2 c_2 c_2 | \Psi \rangle = (2/N_\phi) \langle \Psi | V_1 c^\dagger_1 c^\dagger_2 c_2 c_1 | \Psi \rangle = a. \tag{S14}$$

Once we have Eqs. (S14), we notice that Eq. (S13) reduces to the third term of Eq. (S12). To find the number $a$, we carried out an exact diagonalization on finite systems on the sphere for the singlet $2/3$ state, with the Hamiltonian simply being the bare Coulomb interaction. The results for the singlet $2/3$ state are presented below for up to 14 electrons. Extrapolating to the thermodynamic limit we obtain $a = 0.164$

Furthermore, it is found that $\langle V_0 \rangle \ll \langle V_1 \rangle$ (by three orders of magnitude). Therefore, in our calculations we set the average $\langle \frac{1}{3}, \frac{1}{3} | \hat{V}_0 | \frac{1}{3}, \frac{1}{3} \rangle$ to zero. Since the singlet state has a $U(2)$ symmetry of rotations within the 2D space spanned by the two spinors $|f_{n+1}\rangle$, $|f_{n+2}\rangle$. Thus, the contribution to the energy functional from this subspace must depend only on $P_1 + P_2$, as is verified in Eq. (S12).

In the next sections, we will present phase diagrams for $(1, 2/3, 0, 0)$ and $(1, 1/3, 1/3, 0)$. However, we would like to point out that the phase diagram for $(2/3, 0, 0, 0)$ is identical to that of $(1, 0, 0, 0)$ found in Ref. [5, 7], and that the phase diagram of $(1/3, 1/3, 0, 0)$ is identical to that of $(1, 1, 0, 0)$ found in Ref. [8], for a specific choice of parameters. We will explore these phase diagrams in detail in a forthcoming publication [9]

| $N_\phi$ | $N$ | $\langle V_0\rangle$ | $\langle V_1\rangle$ | $\langle V_0/N\rangle$ | $\langle V_1/N\rangle$ |
|---|---|---|---|---|---|
| 5 | 4 | 0.0007 | 1.4859 | 0.0002 | 0.3715 |
| 8 | 6 | 0.0023 | 2.9583 | 0.0004 | 0.4931 |
| 11 | 8 | 0.0042 | 4.4245 | 0.0005 | 0.5531 |
| 14 | 10 | 0.0064 | 5.9104 | 0.0006 | 0.5910 |
| 17 | 12 | 0.0089 | 7.3871 | 0.0007 | 0.6156 |
| 20 | 14 | 0.0115 | 8.8622 | 0.0008 | 0.6330 |

TABLE S1: The pair-amplitudes in relative angular momentum $m=0,1$ for the exact $\nu=2/3$ spin-singlet Coulomb ground state obtained using exact diagonalization in the spherical geometry for $N$ electrons at flux $N_\phi=3N/2-1$.

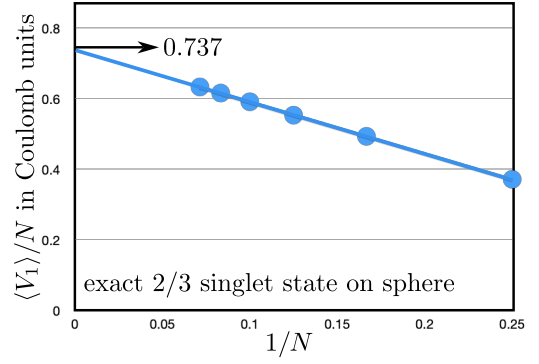

FIG. S1: Thermodynamic extrapolation of the pair-amplitude in relative angular momentum $m=1$ for the $2/3$ spin-singlet Coulomb ground state from finite system results in the spherical geometry.

### SIV. PHASE DIAGRAMS FOR $(1,[2,0]_{-2},0)\equiv(1,\frac{2}{3},0,0)$

Let us briefly consider the USR limit, which was examined variationally at $E_v=0$ by Sodemann and MacDonald [3]. They found five phases: FM : $|K,\uparrow\rangle$, $|K',\uparrow\rangle$, CDW : $|K,\uparrow\rangle$, $|K,\downarrow\rangle$, BO : $|\hat{e}_x,\uparrow\rangle$, $|\hat{e}_x,\downarrow\rangle$, AF : $|K,\uparrow\rangle$, $|K',\downarrow\rangle$, CAF : $|K,\mathbf{s}_a\rangle$, $|K',\mathbf{s}_b\rangle$. In addition to these phases, we find an additional spin-valley entangled phase at $E_v=0$ as shown in Fig. S3a, which we label SVE4 (coral pink). This discrepancy with previous work occurs because the variational states used there (either valley ordered or spin ordered [3]) are too restrictive to find this phase. The spinors that form the ground state in the SVE4 phase are
$\cos\frac{\alpha_1}{2}|\hat{e}_x,\uparrow\rangle-\sin\frac{\alpha_1}{2}|-\hat{e}_x,\downarrow\rangle$, $\cos\frac{\alpha_2}{2}|\hat{e}_x,\downarrow\rangle-\sin\frac{\alpha_2}{2}|-\hat{e}_x,\uparrow\rangle$. This phase has nonzero SVEY but vanishing SVEX at $E_v=0$. It may seem that one could rotate the SVEY order parameter to the SVEX order parameter by a combination of a $U(1)_s$ and a $U(1)_v$ rotation. However, recall that we have already used up the $U(1)_s\otimes U(1)_v$ freedom to make the spinors real. The phase SVE4 also manifests nonvanishing CAF and BO order parameters.

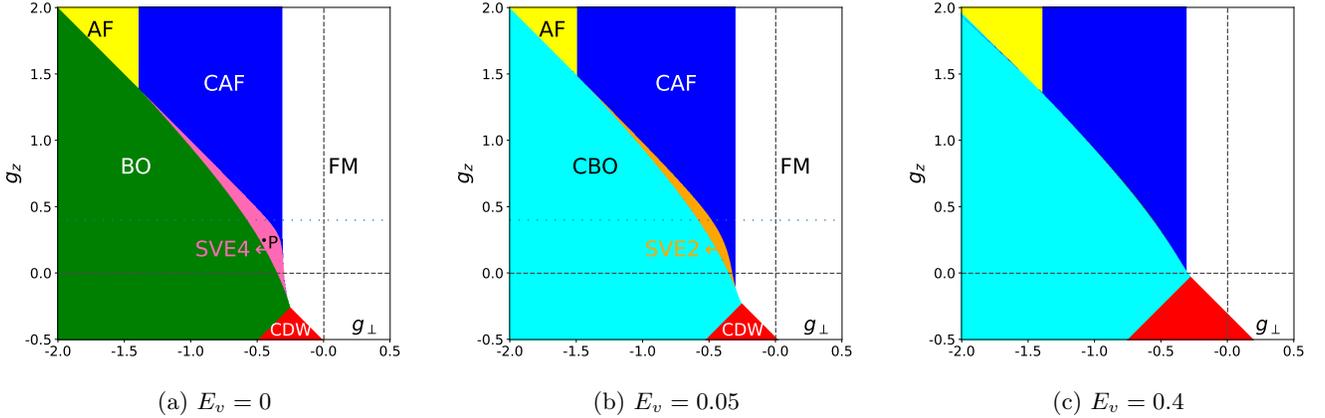

(a) $E_v=0$    (b) $E_v=0.05$    (c) $E_v=0.4$

FIG. S2: Phase diagrams for $(1,2/3,0,0)$ and $E_z=1$ in the USR case. (a) For $E_v=0$ we obtain all the previously obtained phases [3], but also a SVE phase (SVE4, coral pink). (b) At small nonzero $E_v=0.05$, the SVE phase changes character to become SVE2. (c) As $E_v$ increases, eventually, the SVE phase disappears.

The states at $E_v=0$ evolve as $E_v$ increases, as seen in Figs. S2a and S2b. The SVE4 phase changes character and becomes the phase we label SVE2 (orange) as soon as $E_v\neq 0$, the difference being that in SVE2 both SVEX and SVEY are nonvanishing. The SVE2 phase shrinks rapidly as $E_v$ increases and vanishes at some critical $E_v$.

Fig. S3 shows three different views of the order parameter evolution. In Fig. S3a we show a section through the $E_v=0$ phase diagram at $g_z=0.4$ as a function of $g_\perp$. It is clear that the system undergoes a second-order phase transition from the BO phase to the SVE4 phase, and then a first-order transition to the FM phase. Fig. S3b shows

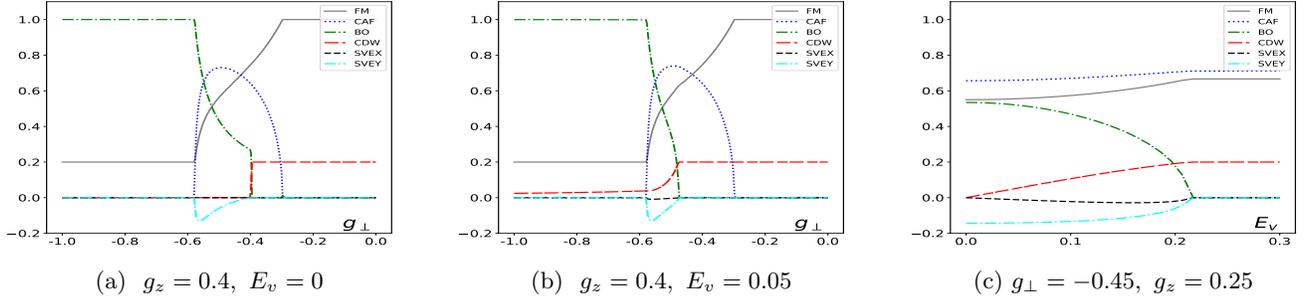

FIG. S3: Different views of order parameter evolution in the USR case for $(1, 2/3, 0, 0)$. (a) A section at $g_z$=0.4 in the phase diagram shown in Fig. S3a. (b) A section at $g_z$=0.4 in the phase diagram shown in Fig. S3b. (c) Order parameters at $g_\perp = -0.45$, $g_z = 0.25$ as functions of $E_v$. It is seen that the SVE order parameters disappear at $E_v \approx 0.23$.

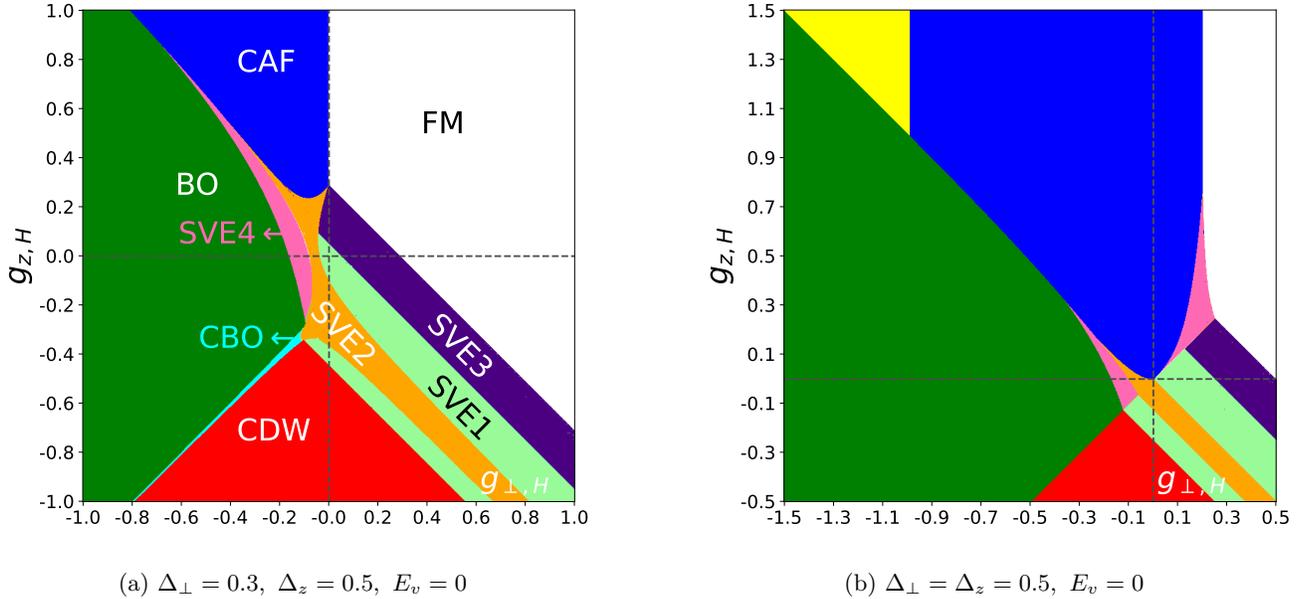

FIG. S4: Phase diagrams with non-USR interactions at $E_z$=1, $E_v$=0 for $(1, 2/3, 0, 0)$. (a) For $\Delta_\perp, \Delta_z$ tiny CBO appears between the BO and CDW phases. The most striking difference from the USR case is the appearance of four SVE phases. (b) The CBO disappears if we fine-tune to $\Delta_\perp = \Delta_z$ but the SVE phases remain.

the same section at $E_v$=0.05. The SVE2 phase has shrunk considerably, and the first-order transition to the FM has become second-order. Fig. S3c shows the evolution of the order parameters with $E_v$ at a specific point in the parameter space $g_\perp$=−0.45, $g_z$=0.25. One sees that as soon as $E_v$ becomes nonzero, the SVEX order parameter also becomes nonzero. At that specific point, the SVE2 phase ceases to be the ground state at around $E_v$=0.23. Now we go beyond USR interactions. When $V_1$ is turned on, several more phases are found, especially at $E_v$=0 and small $E_v$. As seen in Fig. S4a, a canted BO (CBO: $|\mathbf{n}_a, \uparrow\rangle$, $|\mathbf{n}_b, \downarrow\rangle$, in aqua) phase, which is observed at integer fillings only for nonzero $E_v$, now appears between the BO and CDW phases. The appearance of this phase can be qualitatively understood as the result of the unequal fillings of the two spinors because the charge-ordering fields generated by $g_z$ and the nonzero valley polarization of the two spinors do not cancel. We note that the CBO phase appears only when $\Delta_z \neq \Delta_\perp$, as seen from the contrast between Figs. S4b and S4a. In addition, four distinct SVE phases appear in the phase diagram. We have already noted the phases SVE4 and SVE2 in the USR case. The spinors characterizing the new phases are: SVE1 : $|K, \uparrow\rangle$, $\cos(\alpha/2)|K', \uparrow\rangle - \sin(\alpha/2)|K, \downarrow\rangle$ with $\cos \alpha = (2(g_{\perp, H} + g_{z, H} - \Delta_\perp) - E_v + E_z)/\Delta_z$ and SVE3 : $\cos(\alpha/2)|K, \uparrow\rangle - \sin(\alpha/2)|K', \downarrow\rangle$, $|K', \uparrow\rangle$ with $\cos \alpha = (4(g_{\perp, H} + g_{z, H} - \Delta_\perp) + 3E_v + 3E_z)/6\Delta_z$, The difference between the phases SVE1 and SVE3 is in the

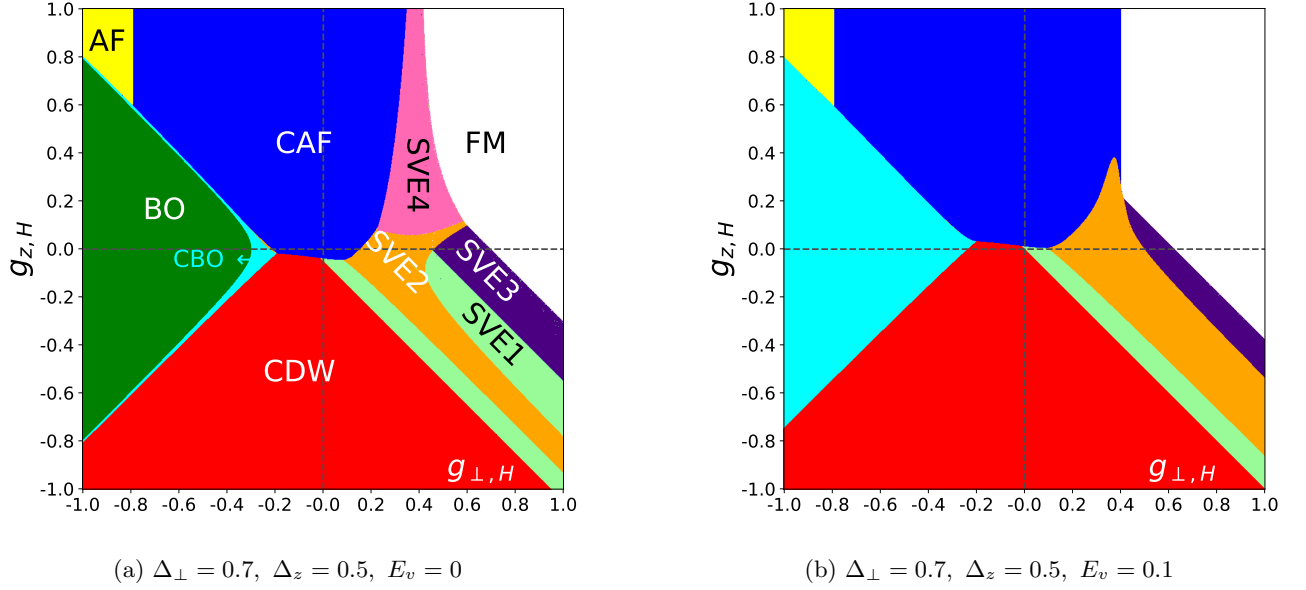

(a) $\Delta_\perp = 0.7$, $\Delta_z = 0.5$, $E_v = 0$

(b) $\Delta_\perp = 0.7$, $\Delta_z = 0.5$, $E_v = 0.1$

FIG. S5: Phase diagrams at $E_z=1$ for $\Delta_\perp > \Delta_z$ for $(1, 2/3, 0, 0)$. (a) A tiny CBO phase appears between the BO and CDW phases at $E_v=0$. This can be understood as a result of the imbalance of the valley ordering energies of the integer and fractional spinors. (b) When a finite $E_v$ is turned on, SVE4 disappears, and the entire BO region becomes CBO, as expected.

relative sign of the order parameters SVEX and SVEY. In phase SVE1, $\langle\tau_x\sigma_x\rangle=\langle\tau_y\sigma_y\rangle$, while in phase SVE3, $\langle\tau_x\sigma_x\rangle=-\langle\tau_y\sigma_y\rangle$.

Fig. S5 shows the evolution of the phase diagram as $E_v$ is turned on. We have already seen in the previous subsection that the phase SVE4 turns into the phase SVE2 upon being subjected to a nonzero $E_v$. This can also be seen in Figs. S5a and S5b. In fact, SVE2 is seen to take over one region of SVE1 as well. Furthermore, the entire BO phase turns into the CBO phase, which is expected.

## SV. PHASE DIAGRAMS FOR $(1, [1,1]_{-2}, 0) \equiv (1, 1/3, 1/3, 0)$

Let us first recall the USR limit for the filling $(1, 1/3, 1/3, 0))$ shown in Fig. S6a, which was first obtained in Ref. [3]. There are two types of bond-ordered phases, a valley ferromagnet (VFM) state with an $SU(2)$ degeneracy of the valley orientation, a FM phase, a CDW phase, a CAF phase, and an SVE phase, which we label SVE3 (SVE1 & SVE2 have been reported in the main text). Here are the spinors of various phases observed in Fig. S6.
FM: $|K,\uparrow\rangle$, $|K',\uparrow\rangle$, $|K',\downarrow\rangle$, AFCDW1: $|K,\uparrow\rangle$, $|K,\downarrow\rangle$, $|K',\uparrow\rangle$, BO1: $|-\hat{e}_x,\uparrow\rangle$, $|\hat{e}_x,\uparrow\rangle$, $|\hat{e}_x,\downarrow\rangle$,
BO2: $|-\hat{e}_x,\uparrow\rangle$, $|\hat{e}_x,\uparrow\rangle$, $|-\hat{e}_x,\downarrow\rangle$, CBO: $|-\mathbf{n}_a,\uparrow\rangle$, $|\mathbf{n}_a,\uparrow\rangle$, $|\mathbf{n}_b,\downarrow\rangle$, VFM: $|\mathbf{n}=\hat{e}_x,\uparrow\rangle$, $|K,\downarrow\rangle$, $|K',\downarrow\rangle$,
CAF: $|K,\mathbf{s}_a\rangle$, $|K,-\mathbf{s}_a\rangle$, $|K',\mathbf{s}_b\rangle$, and SVE3: $|\hat{e}_x,\mathbf{s}_a\rangle$, $|\hat{e}_x,-\mathbf{s}_a\rangle$, $|-\hat{e}_x,\mathbf{s}_b\rangle$. Here $\mathbf{n}_a$, $\mathbf{n}_b$, $\mathbf{s}_a$, $\mathbf{s}_b$ in each phase are determined by minimization.

Going beyond USR interactions, interestingly, when $\Delta_z=\Delta_\perp$, the observed phases are the same as those in the USR case, with slightly different domains. This indicates that there is perhaps some hidden symmetry in the model for $\Delta_z=\Delta_\perp$, which we intend to explore in the near future [9]. However, when $\Delta_z\neq\Delta_\perp$, two new phases appear, a canted bond-ordered phase (CBO, in aqua) and a second SVE phase (SVE2, in light blue). Furthermore, the nature of some of the phases depends on which $\Delta$ is larger. In Fig. S6c, we present the case $\Delta_\perp<\Delta_z$. The valley direction of the VFM is now pinned to be equatorial, making it the BO3 phase. For $\Delta_\perp>\Delta_z$, as seen in Fig. S7a, the valley direction gets pinned to be polar leading to the AFCDW2 phase: $|K,\uparrow\rangle$, $|K,\downarrow\rangle$, $|K',\downarrow\rangle$. Fig. S7b shows how the phase diagram of Fig. S7a evolves at nonzero $E_v$. The CBO and SVE2 phases disappear.

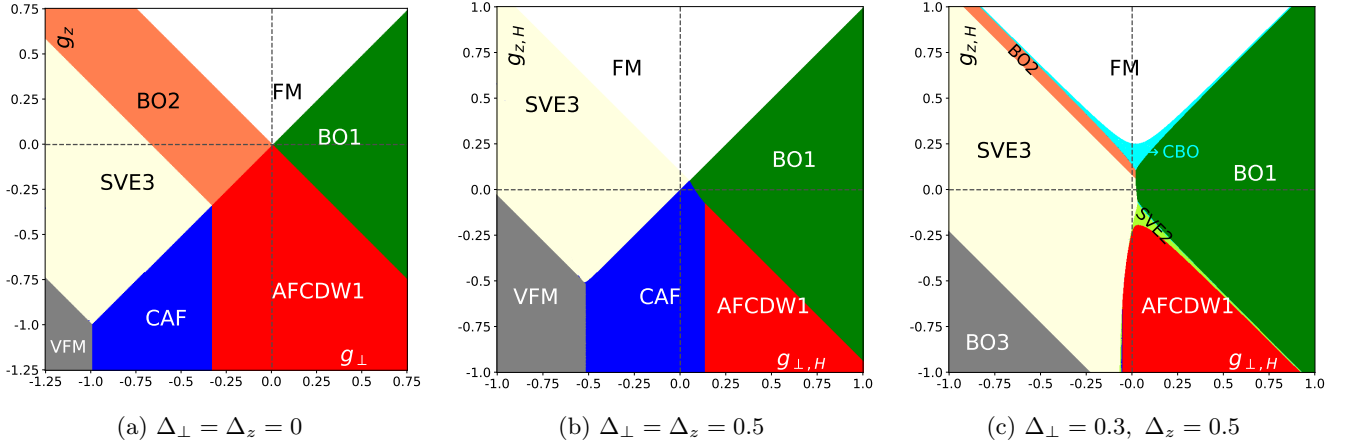

FIG. S6: Phase diagrams for the state $(1, \frac{1}{3}, \frac{1}{3}, 0)$ at $E_z=1$, $E_v=0$. (a) In the USR limit, as described in Ref. [3]. (b) Phase diagram for fine-tuned $\Delta_\perp = \Delta_z$. The nature and overall topology remain the same. The phase BO2 has shrunk to zero. (c) Phase diagram for $\Delta_\perp < \Delta_z$. The VFM changes character to become bond-ordered (BO3), with the **n** vector of the VFM state becoming equatorial. Also, a CBO phase and a SVE phase appear at the boundaries of the old phases.

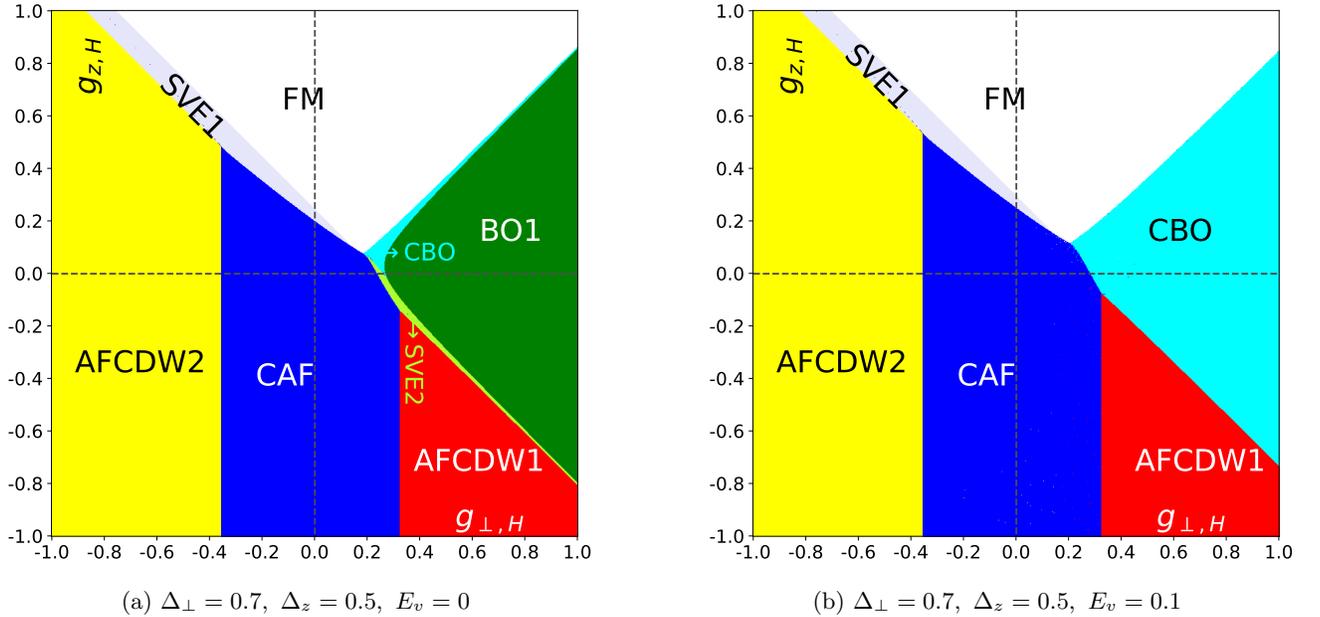

FIG. S7: Phase diagrams of $(1, 1/3, 1/3, 0)$ at $E_z=1$ for $\Delta_\perp > \Delta_z$. (a) At $E_v=0$ the valley ordering vector of the VFM phase in the USR case gets pinned to the polar direction, making this phase the AFCDW2 phase. Further, three new phases that did not occur for the USR case appear: SVE1, SVE2, and CBO. (b) When $E_v$ is turned on, the CBO and SVE2 phases disappear.

## SVI. COULOMB PHASE DIAGRAM OF TWO- AND THREE-COMPONENT STATES

As mentioned in the main text, once one considers the Coulomb interaction, there can be transitions between 2CS and 3CS states driven by the different scaling of the Coulomb and short-range interaction strengths with $B_\perp$, or the dielectric screening $\kappa$. We will carry out this exercise for the states $(1, 2/3, 0, 0)$ and $(1, 1/3, 1/3, 0)$. The Coulomb energy difference between these two states is determined solely by their fractional parts [3], i.e.

$$\delta E^C = E^C_{(1,\frac{2}{3},0,0)} - E^C_{(1,\frac{1}{3},\frac{1}{3},0)} = E^C_{\frac{2}{3}} - E^C_{(\frac{1}{3},\frac{1}{3})}, \tag{S15}$$

where $E^C_{\frac{2}{3}}$ and $E^C_{(\frac{1}{3},\frac{1}{3})}$ are obtained via exact diagonalization[10].

$$E^C_{\frac{2}{3}} = \frac{2}{3} \times (-0.51829)\frac{e^2}{\kappa \ell_B}, \quad E^C_{(\frac{1}{3},\frac{1}{3})} = \frac{2}{3} \times (-0.52704)\frac{e^2}{\kappa \ell_B}, \tag{S16}$$

where $\ell_B = \sqrt{\hbar c/eB_\perp}$ is the magnetic length. The relative dielectric constant of graphene depends on the encapsulating conditions and is typically in the range $4<\kappa<8$. We choose to illustrate the competition between the 2CS and 3CS states at $B=10$ Tesla. At this field, the Zeeman energy is $E_z=(1/2)g\mu_B B=0.58$ meV [$g=2$ for graphene], and the bare Coulomb energy (sans the dielectric constant) is $e^2/\ell_B=182$ meV. We then express all energies in units of $E_z$, i.e., we set $E_z=1$. Therefore, $\delta E^C = E^C_{\frac{2}{3}} - E^C_{(\frac{1}{3},\frac{1}{3})} = 1.84/\kappa$. To decide which state prevails, we compute the total energy difference

$$\Delta E = E^{\mathrm{an}}_{(1,\frac{2}{3})} - E^{\mathrm{an}}_{(1,\frac{1}{3},\frac{1}{3})} + \delta E^C. \tag{S17}$$

In Fig. S8 we show the results for $\kappa=4,6,8$ for the non-USR interactions characterized by $\Delta_\perp=0.7$, $\Delta_z=0.5$. The leftmost column of the two panels shows the phase diagrams of the 2CS and 3CS. In each column of panels, if the energy of the 2CS/3CS state is lower than its competitor in a particular region of the parameter space, it is shown in color, whereas if its energy is higher than its competitor, the region is colored black. As expected, it is seen that for large anisotropic couplings, the anisotropy energy of the 2CS can overwhelm the Coulomb energy advantage of the 3CS. As the screening increases, the Coulomb energy advantage of the 3CS shrinks, and its region of appearance in the phase diagram also shrinks.

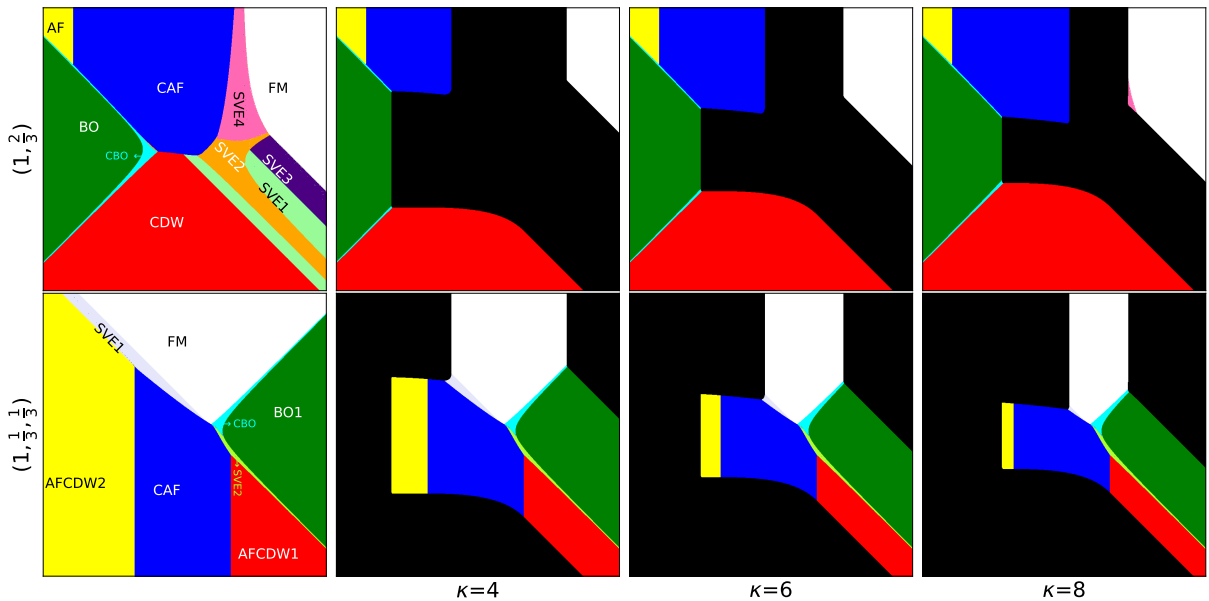

FIG. S8: The competition between the two- and three-component states at $\nu=-2+5/3$, including the Coulomb and residual anisotropy energies for $\Delta_\perp=0.7$, $\Delta_z=0.5$. The leftmost vertical panels are the phase diagrams of the 2CS (top) and 3CS (bottom) states. Each successive set of two vertically aligned panels shows the regions where each type of state has lower energy (in color in the corresponding phase diagram) or higher energy (in black). The three sets of panels are for $\kappa=4,6,8$ respectively.

 

\end{document}